\documentclass[twocolumn,showpacs,preprintnumbers,amsmath,amssymb]{revtex4}

\usepackage{amssymb}
\usepackage{graphicx}
\usepackage{dcolumn}
\usepackage{bm}

\newcommand{\EE}[1]{\mathbb{E}\left[#1\right]}

\begin{document}

\preprint{APS/123-QED}

\title{Uncovering latent singularities from multifractal scaling laws in mixed asymptotic regime. Application to turbulence}

\author{J.F. Muzy}
\email{muzy@univ-corse.fr}
\affiliation{SPE UMR 6134, CNRS, Universit\'e de Corse, 20150 Corte, France}
\author{E. Bacry}
\affiliation{CMAP, Ecole Polytechnique, 91128 Palaiseau, France}
\author{R. Baile}
\affiliation{SPE UMR 6134, CNRS, Universit\'e de Corse, Route des Sanguinaires, Vignola, 20200 Ajaccio, France}
\author{P. Poggi}
\affiliation{SPE UMR 6134, CNRS, Universit\'e de Corse, Route des Sanguinaires, Vignola, 20200 Ajaccio, France}

\date{\today}

\begin{abstract}
In this paper we revisit an idea originally proposed by Mandelbrot
about the possibility to observe ``negative 
dimensions'' in random multifractals. 
For that purpose, we define a new way to study scaling
where the observation scale $\tau$ and the total sample
length $L$ are respectively going to zero and to infinity. 
This ``mixed'' asymptotic regime is parametrized by an exponent
$\chi$ that corresponds to Mandelbrot ``supersampling exponent''.
In order to study the scaling
exponents in the mixed regime, we use 
a formalism introduced in the context of the physics 
of disordered systems relying upon traveling wave solutions
of some non-linear iteration equation. Within our approach, we show that for 
random multiplicative cascade models, 
the parameter $\chi$ can be interpreted 
as a negative dimension and, as anticipated by Mandelbrot, 
allows one to uncover the ``hidden'' 
negative part of the singularity spectrum, corresponding 
to ``latent'' singularities. 
We illustrate our purpose on synthetic cascade models.
When applied to turbulence
data, this formalism allows us to distinguish 
two popular phenomenological models of dissipation intermittency: We show that
the mixed scaling exponents agree with a log-normal model and not with log-Poisson
statistics.
\end{abstract}

\pacs{Valid PACS appear here}
\maketitle
Multifractal processes are random functions (or measures) that possess 
non trivial scaling properties.
They are widely used models in many areas
of applied and fundamental fields. Well known examples 
are turbulence, internet traffic, rainfall distributions or finance. 
For the sake of simplicity we will consider only non-decreasing multifractal 
processes (which increments define a multifractal measure) 
denoted hereafter $M(t)$.
In the sequel $M(I)$ will stand for the measure of the interval $I$,
$M(I) = \int_I dM$ and $M(t,\ell) = M([t,t+\ell])$.
Multifractals are characterized by the scaling of the partition
functions: If one covers the overall sample interval of length $L$ with
$L/{\ell}$ disjoint intervals of size $\ell$, $\{I_\ell(i) \}_{i=1\ldots L/\ell}$,
one usually defines the order $q$ partition function which scaling 
behavior defines the exponent $\tau_0(q)$:
\begin{equation}
\label{pf1}
  Z(q,\ell) = \sum_{i=1}^{L/\ell} M\left[I_\ell(i)\right]^q \operatornamewithlimits{\sim}_{\ell \rightarrow
   0} \ell^{\tau_0(q)}
\end{equation}
where the limit $\ell \rightarrow 0$ simply means that $\ell/T\rightarrow 0$,
$T$ being the correlation scale usually referred to as the integral scale.
When $\tau_0(q)$ is a (concave) nonlinear function of $q$, the 
measure $M(t)$ is said to be multifractal or intermittent. Within the multifractal
formalism introduced by Parisi and Frisch (see e.g. \cite{BF}), the non-linearity of
$\tau_0(q)$ is interpreted in terms of fluctuations of pointwise singularity exponents
of the measure. Indeed, according to this formalism, $\tau_0(q)$ is obtained as 
the Legendre transform of the singularity spectrum $f_0(\alpha)$ that gives the
(Haussdorf) dimension of the sets of points $t$ of singularity $\alpha$ 
($M(t,\ell) \sim \ell^\alpha$). Therefore $q$ can be interpreted as a 
value of the derivative of $f_0(\alpha)$ and conversely
$\alpha$ is a value of the derivative of $\tau_0(q)$. 
Multifractality is also closely related to the notion of stochastic self-similarity:
The measure $M(t)$ is self-similar in a stochastic sense if, for all $s < 1$,
$M(st) \operatornamewithlimits{=}_{law} s W_s M(t)$
where $W_s$ is a positive random weight independent of $M$.
It can be easily shown that this stochastic equality implies that
the expected value of $Z(q,\ell)$ (i.e., the order $q$ moment of the measure) 
behaves as a power law, i.e., $\EE{Z(q,\ell)} = C_q \ell^{\tau(q)}$
where $\tau(q)$ is nothing but the cumulant generating function 
of $\ln W_s$. If one uses the terminology introduced in physics of disorderded
systems, the exponent $\tau(q) \sim \ln \EE{Z(q,\ell)}$ is defined from 
an ``annealed'' averaging while $\tau_0(q) \sim \EE{\ln Z(q,\ell)}$ 
is the analog of a free energy computed as a ``quenched'' average. 
As it will be discussed below, these two functions can be different for large values of $q$.

The paradigm of self-similar measures
are random multiplicative cascades (for the sake of simplicity we will exclusively focus, in this paper, on discrete cascades but
all our results can be easily extended to recent continuous cascade
constructions \cite{MuDeBa00,BaMan02,BaMu03}) 
originally introduced by the russian school for modelling the energy 
cascade in fully developed turbulence and to which  
a lot of mathematical studies have been devoted \cite{Man74a,KaPe76,Mol96,Liu02}.
Let us summary the main properties of these constructions.
The integral scale from which the cascading process ``starts'' is denoted as $T$. 
A dyadic discrete cascade is build as follows: A measure is
uniformely spread over the starting interval $[0,T]$ and one
splits this interval in two equal parts: On each 
part, the density is multiplied by (positive) i.i.d. random factors $W$ such
that $\EE{W} = 1$.
Each of the two sub-intervals is again cut in two equal parts and
the process is repeated infinitely. It is convenient to introduce the random
variable $\omega = \ln(W)$ which probability density function will be denoted as $g(x)$.
Thus $g(x)$ is Gaussian or Poisson for respectiveley log-Normal and log-Poisson cascades.
The stochastic self-similarity property
of the limit measure $M$ associated with previous iterative construction 
can be directly proven and it is easy to show that
$\tau(q)$ is related to the cumulant generating function of $\omega$:
\begin{equation}
\label{deftau}
  \tau(q) = q-\ln_2(\EE{e^{q\omega}})-1
\end{equation}
Moreover, as first established by Molchan \cite{Mol96,Mol97},  
the multifractal formalism holds for random cascades and one has
the following relationship between $\tau_0(q)$ and $\tau(q)$, for $q>0$, (note that a similar relationship holds for negative values of $q$ \cite{Mol96}):
\begin{equation} 
\label{tau0}
  \tau_0(q)  = \tau(q) \; \; \mbox{if} \; \; q \leq q_0 \; \; \mbox{and}  \; \; \tau_0(q)  =  \alpha_0 q \; \; \mbox{if} \; \; q  > q_0 
\end{equation}
where $q_0$ is the value of $q$ corresponding to minimum value of $f_0(\alpha)$ (in general $f_0(\alpha_0) = 0$).
This discrepency between the annealed and quenched spectra has been
extensively studied on a numerical ground in ref. \cite{LAC04} and was referred to as the ``linearization effect''. 
It has notably been observed to be independent of the nature of the cascade and of the overall 
length $L$ of the sample. As it will be emphasized below this effect has not been properly taken
into account in the literature of turbulence (see e.g. \cite{VanDe99}).

One of the goals of this paper is to recover the linearization effect and to establish how it can be somehow controlled.
For that purpose, we will consider the scaling of partition function \eqref{pf1} in some ``mixed'' asymptotic  regime where, as the 
resolution becomes smaller, the total length of the sample is inscreased.
Let us introduce some useful notations: We call $\ell$ the scale of observations (it can be the sampling scale or a multiple of it), 
$T$ the integral scale and $L$ the total sample length. 
$N_{\ell}$ will refer to the number of samples per integral scale and $N_T$ the number of 
integral scales. We have obviously $N_\ell  =  T \ell^{-1}$ and $N_T  =  L T^{-1}$ and $N = N_{l}N_{T} = L/\ell$ is the total number of samples.
If $T$ is fixed, the limit $N \rightarrow +\infty$  can
be conveniently controlled using an additional exponent $\chi \geq 0$ as
$ N_T  \sim N_{\ell}^{\chi}$.
Let us mention that such an exponent has been already introduced by B.B. Mandelbrot as an ``embedding dimension'' \cite{Man90,ChaSree91} 
in order to discuss the concept of negative dimension and latent singularities (see below).
Within this framework, the definition \eqref{pf1} of the partition function depends on $\chi$ and
allows us to define a new exponent $\tau_\chi(q)$ as follows: 
\begin{equation}
\label{pf2}
  Z(q,\ell,\chi) = \sum_{i=1}^{L \ell^{-1}} M\left[I_\ell(i)\right]^q \operatornamewithlimits{\sim}_{\ell \rightarrow
   0} \ell^{\tau_\chi(q)-\chi}
\end{equation} The two ``extreme'' cases are : (i)
the $\chi = 0$ case which corresponds to a fixed
number of integral scales $N_{T}$ while  $l \rightarrow 0$ and (ii) the $\chi = +\infty$ case which 
corresponds to a fixed  observation scale $\ell$  while $N_{T}\rightarrow +\infty$. It results that for $\chi=0$ one recovers former definition \eqref{pf1} of $\tau_0(q)$ 
while $\tau_\infty(q) = \tau(q)$ as defined in \eqref{deftau}. In that respect $\chi$ allows us
to interpolate between quenched and annealed situations. 
In order to compute $\tau_\chi(q)$ one needs to study the behavior of the probability law of $Z(q,\ell,\chi)$ \cite{inprephof}.
For that purpose, along the same line as in references \cite{DerSpo88,CarLeDou01}, let 
us study its Laplace transform: $G(s,\ell) = \EE{e^{-e^{-qs} Z(q,\ell,\chi)}}$. 
Let $r,p$ two integers and let us define the iteration $m \rightarrow m+1$, $\ell \rightarrow 2^{-p} \ell$ and $L \rightarrow 2^{r} L$.
In other words, at each iteration step, the resolution is divided by
$2^p$ while the number of independent integral scales is multiplied by
$2^r$. One thus have $N_T = 2^{rm}$ and 
$N_\ell = 2^{pm}$ that corresponds to $\chi = r/p$.
Within this parametrization, $G(s,\ell)$ will be denoted as $G(s,m,p,r)$
and $G(s,m,1,0)$ will be denoted as $H(s,m)$. If $M$ is a random cascade
as defined previously, its self-similarity allows 
one to prove that $G(s,m,p,r)$ can be 
written as \cite{inprep}:
\begin{equation}
\label{iter2}
  G(s,m,p,r) =  \left(H(s,pm) \right)^{2^{rm}} \; ,
\end{equation}
where $H(s,m)$ satisfies the following recursion
\begin{equation} 
\label{iter1}
   H(s,m+1) = \left[ H(s,m) \ast g(s+\ln 2) \right]^2
\end{equation}
where $g(x)$ is the pdf of $\omega$ the logarithm of cascade weights and $\ast$
stands for the convolution product.

\begin{figure}
\includegraphics[height=5.5cm]{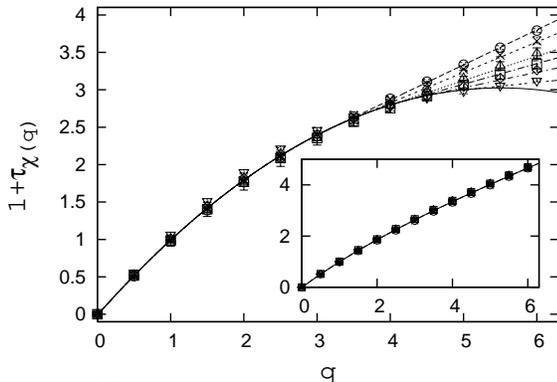}
\vskip -0.7cm
\caption{\label{fig1} $\tau_\chi(q)$ estimated for log-normal random cascade for 
$\chi = 0$ ($\circ$), 0.2 ($\times$), 0.5 ($\bigtriangleup$), 0.7 ($\Box$), 1 ($\Diamond$), 1.5 ($\bigtriangledown$). For each value of $\chi$, we have chosen $T = 4096$,
$\lambda^2 = 0.2$ and $N_T$ depends on $\chi$ such that $N_T N_\ell$ is fixed. One clearly sees that, as the value of $\chi$
increases, the value of $q_\chi$ increases, the slope $\alpha_\chi$ of $\tau_\chi(q)$ decreases and
$\tau_\chi(q)$ becomes closer to the annealed spetrum $\tau(q)$ (solid line). The dashed lines represent
the analytical expressions of $\tau_\chi$ derived from Eqs. \eqref{tauchi1} and \eqref{tauchi2}. 
In the inset the same estimations have been performed for a log-Poisson cascade. As expected, in that case $\tau_\chi(q)$ 
weakly depends on $\chi$ and rapidly converges towards $\tau(q)$ represented by the solid line (see text)}
\end{figure}

It is easy to see that Eq. \eqref{iter1} as two uniform ``stationnary' solutions
$G(s,n) = 0$ and $G(s,n)=1$, the first one being stable while the latter is
(linearly) unstable. 
The initial condition connects the stable state $H(-\infty,0)=0$
to the unstable one $H(\infty,0) = 1$ and, 
as shown in \cite{DerSpo88,Brunet,DeMaj01}, this kind of
non-linear equation admits traveling wave solutions $H(s,n) = H_0(s-vn)$ 
where $v$ is the front velocity. This velocity can be computed
by studying the linearized version of Eqs. \eqref{iter2} 
and \eqref{iter1} around the unstable state. 
After a little algebra \cite{inprep}, one can show
that the solutions of \eqref{iter2} are traveling fronts that can be written, 
when $s \rightarrow +\infty$, as $G(s,m,p,r) = (1-Ce^{-\gamma(x-v(\gamma)pm)})^{2^{rm}}$, provided $v(\gamma)$ satisfies the dispersion relation: 
$v(\gamma) = \frac{\ln(2) \left(r-p\tau(\gamma)\right)}{\gamma}$ where 
$\tau(\gamma)$ is defined by Eq. \eqref{deftau}. One can reproduce the same kind of analysis as in 
refs. \cite{Brunet,BruDer97,MajKra01} 
and show that a standard Aronson-Weinberger stability criterium can be used
to compute the selected velocity and $\gamma$ values:
Let $q_\chi$ be the unique positive $\gamma$ value such that $v(q_\chi) = \min_{\gamma>0} v(\gamma)$.
The selected velocity $v(\gamma)$ actually corresponds to the $\gamma$ value equal
to the exponential decreasing rate of the initial condition provided it is greater than $q_{\chi}$. 
But the initial condition $H(s,0)$ is precisely given by the
Laplace transform of the unconditional law of $M[0,1]^q$ and 
if $\EE{M^q} = M_q < +\infty$ (such a condition is well known to be satisfied provided $\tau(q) > 0$)
then $H(s,0) \operatornamewithlimits{\sim}_{s \rightarrow \infty} 1-M_q e^{-qs}$.
Therefore, the selected velocity is simply $v = \frac{\ln(2)\left[r-p\tau(q))\right]}{q}$ if $q < q_{\chi}$
and $v = v(q_{\chi})$ otherwise. Thanks to the fact that $m$ is related to the resolution scale by $\ln(\ell)= -m p \ln(2)$, 
and using $\chi = r/p$, the velocity, measured as respect to $\ln(\ell)$, 
finally becomes $v(q)= \frac{\tau(q)-\chi}{q}$ if $q < q_{\chi}$ and $v(q)= \max_{q>0} \frac{\tau(q)-\chi}{q}$ 
otherwise. If one denotes $f(\alpha)$ the legendre transform of $\tau(q)$, the value $q_{\chi}$ for wich the maximum value 
of $\frac{\tau(q)-\chi}{q}$ is reached, satisfies  $q_{\chi}  =  f'(\alpha_{\chi})$ with
\begin{equation}
\label{defalphastar}
  f(\alpha_{\chi}) =  -\chi \; \mbox{and} \; \alpha_\chi = \tau'(q_\chi) = v(q_\chi)    
\end{equation}
Since positive values of $f(\alpha)$ correspond to a fractal dimension, the $\chi$ value
can be seen as a kind of ``negative dimension''. 
In order to solve the initial problem one can reproduce the analysis of \cite{DerSpo88}
to show that the front velocity is directly related to the mean value of $\ln Z(q,\ell,\chi)$
as $q^{-1} \EE{\ln Z(q,\ell,\chi)}  \sim  v(q) \ln(\ell)$.
Since in the ``moving frame'', the probability distribution of $\ln Z(q,\ell,\chi)$ converges,
it results that the fluctuations of $\ln Z(q,\ell,\chi)/\ln(\ell)$ vanish. This notably implies
that $ q^{-1} \frac{\ln Z(q,\ell,\chi)}{\ln \ell} \rightarrow v(q)$, where the convergence is in probability. According to definition \eqref{pf2}, this is equivalent to say that $\tau_\chi(q)$ is non random and 
equals to $v(q,\chi)+\chi$, i.e.:
\begin{eqnarray}
\label{tauchi1}
   \tau_\chi(q) & = & \tau(q) \; \; \mbox{if} \; \; q < q_{\chi} \\
\label{tauchi2}
   \tau_\chi(q) & = & q \alpha_{\chi} \; \; \mbox{otherwise}
\end{eqnarray}
where $\alpha_{\chi}$ is defined in \eqref{defalphastar}. Let us note that a rigourous proof of Eqs \eqref{tauchi1} and \eqref{tauchi2} will be provided in \cite{inprephof}.
In the case $\chi = 0$ one recovers standard linearization effect (Eq. \eqref{tau0})
which has been generalized to any value of $\chi$ in the mixed asymptotic regime.
As $\chi$ increases so does $q_{\chi}$ and $\tau_\chi(q)$ continuously converges towards
$\tau(q)$. Singularities $\alpha<\alpha_0$ have been qualified by Mandelbrot as ``latent'' 
because they are only observable for large enough values of the ''supersampling'' exponent $\chi$ \cite{Man90}. 

\begin{figure}
\includegraphics[height=5.5cm]{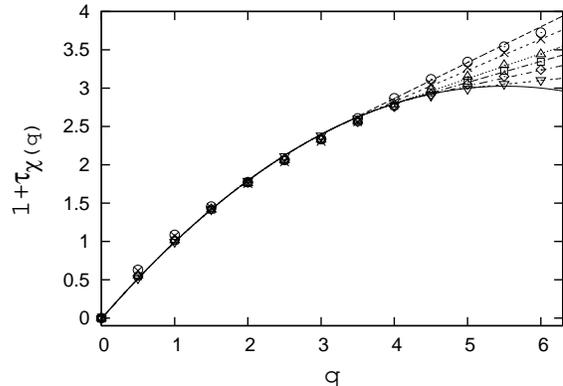}
\vskip -0.7cm
\caption{\label{fig2} Same plot as in Fig. \ref{fig1} but for the dissipation field $\epsilon(x)$
estimated from a high Reynolds number jet turbulence experiment. The symbols have the same meaning
as in Fig. \ref{fig1}. As for the log-Normal synthetic cascade
one clearly observes, as $\chi$ increases, a continuous evolution of $\tau_\chi(q)$ 
towards the analytical parabola corrresponding to a log-Normal model with $\lambda^2 = 0.2$ (solid line). 
Such an effet is not observed for a log-Poisson cascade.}
\end{figure}

In order to illustrate our results, we have performed several numerical simulations on both
continuous and discrete cascades which statistics are log-normal (LN) and log-Poisson (LP). In the log-normal
case one has $\tau(q)= q(1+\lambda^2/2)-\lambda^2 q^2/2-1$ and therefore by solving Eq \eqref{defalphastar} one gets
$\alpha_{\chi} = 1+\lambda^2/2-\lambda \sqrt{2({+\chi})}$ and $q_{\chi} = \lambda^{-1}\sqrt{2(1+\chi)}$.
If Fig. \ref{fig1}, we have plotted, for each value of $\chi = 0,0.2,0.5,0.7,1,1.5$, $\zeta_\chi(q)=1+\tau_\chi(q)$
as a function of $q$ estimated from the scaling of partition 
function $Z(q,\ell,\chi)$ over a wide enough range of scales. The log-normal cascade samples are generated
from discrete cascades with $T = 4096$, $\lambda^2 = 0.2$ and, for each value of $\chi$,  $N_T$ is chosen  such that, 
at the smallest scale, the number of sampling points is almost constant.
As expected, one sees that, when the value of $\chi$ increases, $q_\chi$ increases, $\alpha_{\chi}$ (the
slope of $\tau_\chi(q)$ for $q > q_{\chi}$) decreases and one has $\tau_\chi(q)= \tau(q)$ over a wider range of $q$. 
The same kind of simulations have been performed on log-Poisson synthetic cascades for
which $\tau(q) = q(1+\lambda^2(e^\delta-1)/\delta^2)+\lambda^2(1-e^{q\delta}/\delta^2)-1$. In that case, it is easy
to show that since $\tau(q)$ has an asymptote when $q \rightarrow +\infty$
$\tau'(q) \geq 1+\lambda^2(e^{\delta}-1)/\delta^2)$ and $f(\alpha) \geq -\lambda^2/\delta^2$. Therefore,
for all $\chi \geq \lambda^2/\delta^2$, one has $\tau_\chi(q) = \tau(q)$. If one chooses the log-Poisson 
parameter values usually considered to model energy dissipation in turbulence \cite{BF,SheLev94}, i.e., $\lambda^2 = 0.2$ and
$\delta = \ln(2/3)$, one sees that $\tau_\chi(q)$ rapidly converges towards $\tau(q)$ and no longer varies for $\chi \geq 0.2$. This is illustrated in the inset of Fig. \ref{fig1} where all the estimated 
$\tau_\chi(q)$ for $\chi = 0,0.2,0.5,0.7,1,1.5$ are close or equal to the annealed spectrum $\tau(q)$
represented by the solid line. 

\begin{figure}
\includegraphics[height=5.5cm]{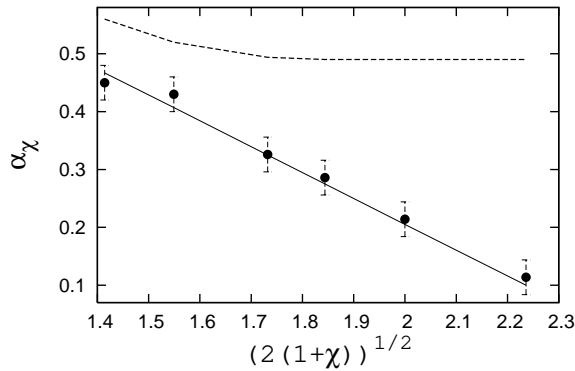}
\vskip -0.7cm
\caption{\label{fig3} Plot of $\alpha_\chi$, the slope of $\tau_\chi(q)$, for $q \geq q_{\chi}$ 
estimated for the turbulent dissipation field 
as a function of $\sqrt{1+\chi}$. The straight solid line represents the log-Normal prediction while
the dashed curve is the log-Poisson prediction. Error bars have been estimated using Monte-Carlo 
trials of the Log-normal cascade. This representation allows us to distinguish
clearly the two models of energy dissipation intermittency.}
\end{figure}

According to our analysis and from numerical experiments reported in Fig. \ref{fig1}, one sees
that despite the fact that $\tau_\chi(q)$ associated with log-normal and log-Poisson are very similar for small values
of $q$ and small values of $\chi$, the situation is very different for large values of $\chi$ and $q$. 
Since both models are traditionnally used to describe the spatial flucuations of energy dissipation in fully
developed turbulence \cite{BF,SheLev94}, we naturally reproduced previous statistical analysis using experimental
data of turbulence. The data have been recorded by the group of B. Castaing in Grenoble in a low temperature gazeous Helium 
jet experiment which Taylor scale based Reynolds number is $R_\lambda = 929$ \cite{chan00}.
Assuming the validity of Taylor hypothesis and isotropy of the flow, a proxy of the dissipation
field $\epsilon(x) \simeq (\partial v/\partial t)^2$ is build from the temporal longitudinal velocity signal.
The overall sample is such that $N_T \simeq 2500$. The estimated values of $\tau_\chi(q)$ are reported in Fig. \ref{fig2}.
The similarity with the results of log-normal synthetic cascades is striking: As in Fig. \ref{fig1}, when $\chi$ increases,
the estimated $\tau_\chi(q)$ converges to the parabolic $\tau(q)$. In Fig. \ref{fig3}, we have plotted 
the values of the asymptotic slope of $\tau_\chi(q)$, $\alpha_\chi$, as a function of $\sqrt{1+\chi}$: one clearly sees that
the data perfectly match the linear log-Normal expression (solid line) which is very different from the log-Poisson prediction (dashed line).

To summarize, we have shown in this paper, using traveling wave solutions of cascade non-linear
iteration equation, that quenched and annealed averaged partitions function have different 
behavior for values of $q$ larger than a critical value $q_\chi$, analog of the glass transition
temperature in spin glass systems. This difference can be controlled using some ''supersampling''
exponent $\chi$ which defines an asymptotic limit that mixes small scales and large number of samples regimes.
By analyzing the value of the multifractal scaling exponents for various values of $\chi$ one can
distinguish between different cascade models. In the context of the modelling of
energy dissipation intermittency in fully developed turbulence, we have provided evidences supporting log-Normal 
statistics against log-Poisson statistics.

\begin{acknowledgments}
\vskip -0.5cm
We thank B. Chabaud and B. Castaing for the permission to
use their turbulence experimental data.
\end{acknowledgments}

\end{document}